\pdfoutput=1
\documentclass[12pt]{iopart}

\usepackage{graphicx}
\usepackage{dcolumn}
\usepackage{caption}

\expandafter\let\csname equation*\endcsname\relax
\expandafter\let\csname endequation*\endcsname\relax 
\usepackage{amsmath}

\usepackage{amssymb}
\usepackage{hyperref}
\usepackage{url}
\usepackage{xspace}

\newcommand {\rootsNN}  	{\ensuremath{\sqrt{s_{_{NN}}}}}
\newcommand {\roots}    	{\ensuremath{\sqrt{s}}}
\newcommand{\GeVc}			{\ensuremath{{\,\text{Ge\hspace{-.08em}V\hspace{-0.16em}/\hspace{-0.08em}}c}}\xspace}
\newcommand{\ptt}			{\ensuremath{p_{\mathrm{T}}}\xspace}
\newcommand {\ptttrg}       {\ensuremath{p_\mathrm{T}^{\mathrm{trig}}}}
\newcommand {\pttass}       {\ensuremath{p_\mathrm{T}^{\mathrm{assoc}}}}

\begin{document}
\title[Ridge correlation structure in high multiplicity pp collisions with CMS]{Ridge correlation structure in high multiplicity pp collisions with CMS}
\author{Dragos Velicanu for the CMS collaboration\footnote{For the full list of
CMS authors and acknowledgments, see appendix ``Collaborations''.}}
\address{
\vspace{2mm}
{\scriptsize
Massachusetts Institute of Technology, 77 Mass Ave, Cambridge, MA 02139-4307, USA\\
}}
\ead{velicanu@mit.edu}
\begin{abstract}

Results on two-particle angular correlations are presented in proton-proton collisions at center of mass energies of 7 TeV, over a broad range of pseudorapidity and azimuthal angle. In very high multiplicity events at 7 TeV, a pronounced structure emerges in the two-dimensional correlation function for particle pairs with intermediate $p_{\mathrm{T}}$ of 1--3 GeV/c, in the kinematic region $2.0 < |\Delta\eta| < 4.8$ and small $\Delta\phi$. This structure, which has not been observed in pp collisions before, is similar to what is known as the "ridge" in heavy ion collisions. It is not predicted by commonly used proton-proton Monte Carlo models and is not seen in lower multiplicity pp collisions. Updated studies of this new effect as a function of particle transverse momentum, rapidity and event characteristics are shown.

\end{abstract}

\vspace{-0.4cm}

Long-range, near-side ($\Delta\phi \approx 0$) ridge-like 
azimuthal correlations for $2.0$$<$$|\Delta\eta|$$<$$4.8$ have 
recently been observed for the first time in high multiplicity pp 
collisions at \roots\ = 7~TeV~\cite{Khachatryan:2010gv}. The novel 
structure resembles similar features observed in relativistic heavy-ion experiments.
This striking feature is most evident in the intermediate 
transverse momentum range of both $1$$<$$\ptttrg$$<$$3$ GeV/c and 
$1$$<$$\pttass$$<$$3$ GeV/c. A steep increase of the near-side 
associated yield with multiplicity has been found in the data.

Following up the first 
observation of the ridge correlation 
structure in high multiplicity pp collisions at \roots\ = 7~TeV, 
new results are presented in this paper to study the detailed
event multiplicity, transverse momentum and pseudorapidity 
gap ($\Delta\eta$) dependence of the ridge effect 
using the full statistics data collected in 2010. 
With the nearly 4$\pi$ solid-angle acceptance of the 
silicon tracker and dedicated high multiplicity
high-level trigger (HLT) setup, the CMS experiment has a unique
capability in studying this novel effect~\cite{JINST}.

The same analysis procedure, as used in the ridge measurement of
the central heavy-ion collisions~\cite{ref:HIN-11-001-PAS}, is
applied in order to make direct comparison to the heavy-ion results, for full details see~\cite{ref:HIN-11-006-PAS}.
The pp data used in this extended analysis are collected under
almost the same condition as those used in the publication of the first pp
ridge observation~\cite{Khachatryan:2010gv} with about a factor of 2 
increase in statistics, 660K $N \geq 110$ events. 

\begin{figure}[thb]
  \begin{center}
    \includegraphics[width=0.49\linewidth]{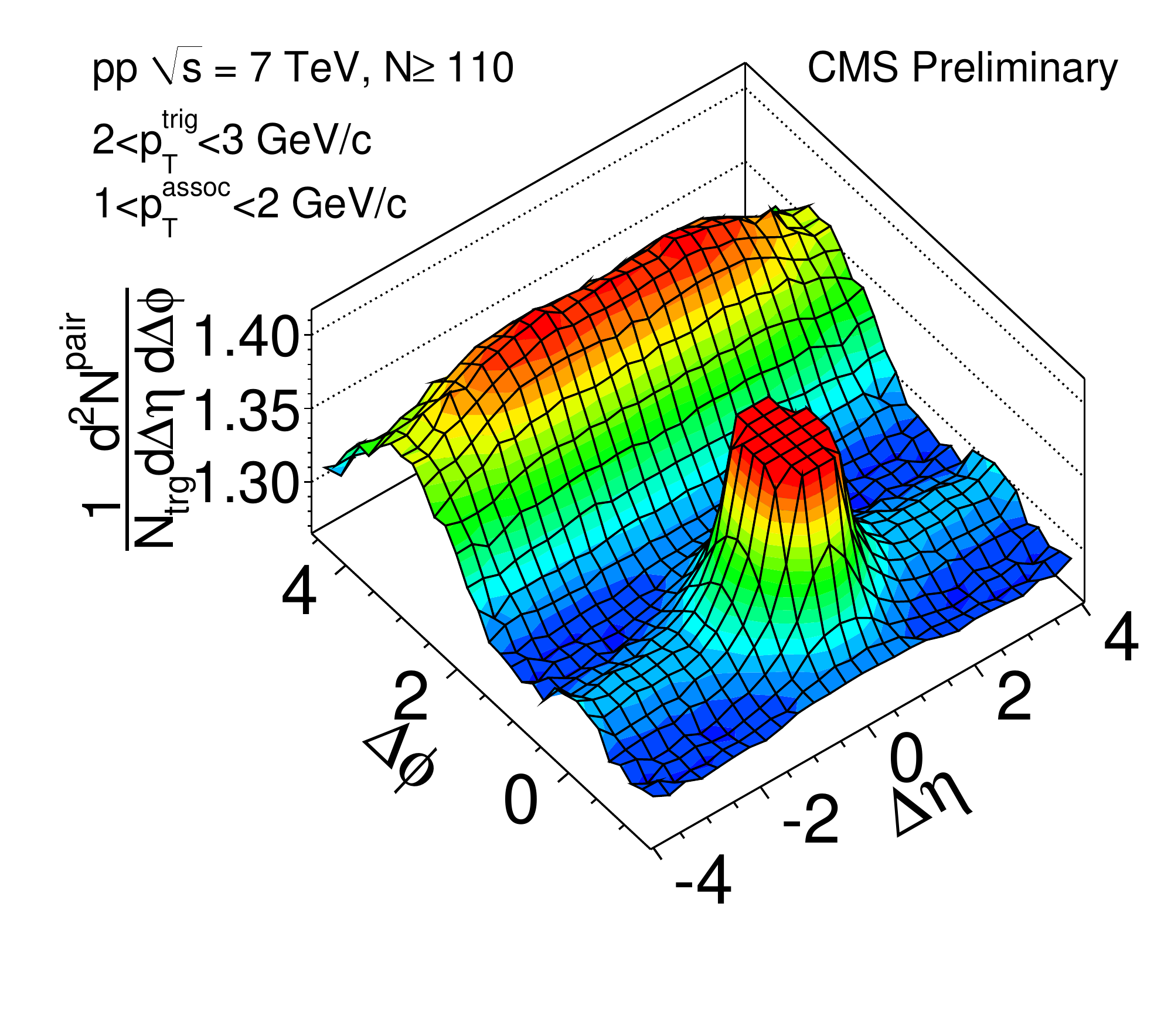}
    \includegraphics[width=0.49\linewidth]{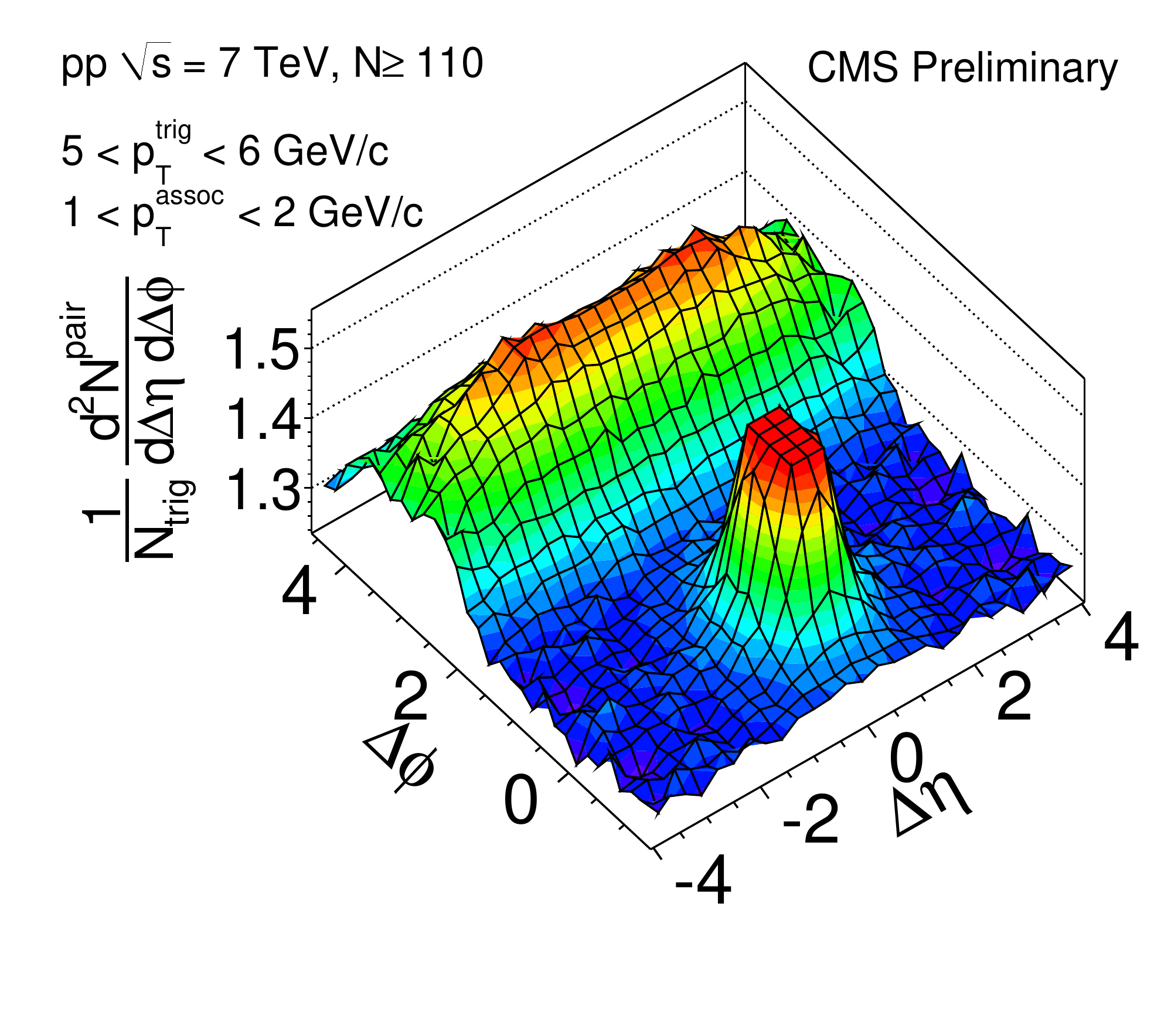}
    \caption{
         Two-dimensional (2-D) per-trigger-particle associated yield of charged hadrons
         as a function of $\Delta\eta$ and $\Delta\phi$
         with jet peak cutoff for better demonstration of the ridge from 
         high multiplicity ($N \geq 110$) pp collisions at \roots\ = 7~TeV, for 
         (a) 
         $2 $$<$$ \ptttrg $$<$$ 3$ GeV/c and $1 $$<$$ \pttass $$<$$ 2$ GeV/c and
         (b) $5 $$<$$ \ptttrg $$<$$ 6$ GeV/c and $1 $$<$$ \pttass $$<$$ 2$ GeV/c 
         }
    \label{fig:corr2D_N110_20110420}
  \end{center}
\end{figure}

The per-trigger-particle associated yield distribution of 
charged hadrons as a function of $\Delta\eta$ and $\Delta\phi$ 
in high multiplicity ($N \geq 110$) pp collisions
at \roots\ = 7~TeV with trigger particles 
with $2$$<$$\ptttrg$$<$$3 \GeVc$ and 
associated particles with $1$$<$$\pttass$$<$$2 \GeVc$ is shown in 
Fig.~\ref{fig:corr2D_N110_20110420} obtained with the full statistics data in 2010. 
The ridge-like structure is clearly visible at $\Delta\phi \approx 0$ 
extending to $|\Delta\eta|$ of at least 4 units as previously observed in
Ref.~\cite{Khachatryan:2010gv}. However, at higher \ptttrg\ of 5--6\GeVc 
as presented in Fig.~\ref{fig:corr2D_N110_20110420}, the ridge almost disappears.
The absolute values of $\Delta\eta$ and
$\Delta\phi$ are used in the analysis, thus the resulting distributions
are symmetric about ($\Delta\eta$,$\Delta\phi$)=(0,0) by construction.

In order to fully explore the detailed properties of both short-range jet-like 
correlations and long-range ridge-like structure, especially its dependence 
on event multiplicity, transverse momentum and $|\Delta\eta|$,
the associated yield distributions are obtained in eight bins ($2 \leq N < 35$, 
$35 \leq N < 45$, $45 \leq N < 60$,
$60 \leq N < 90$, $N \geq 90$,
$N \geq 110$, $N \geq 130$,
$N \geq 150$) of charged particle
multiplicity and six bins (0.1--1, 1--2, 2--3, 3--4, 4--5 and 5--6$ \GeVc$) 
of particle transverse momentum. The 1-D $\Delta\phi$ azimuthal correlation 
functions are calculated by integrating over the $0.0<|\Delta\eta|<1.0$ and 
$2.0<|\Delta\eta|<4.0$ region, defined as the jet region and ridge region, respectively. 

\begin{figure}[thb]
  \begin{center}
    \hspace{-0.8cm} 
    \includegraphics[width=0.6\textwidth]{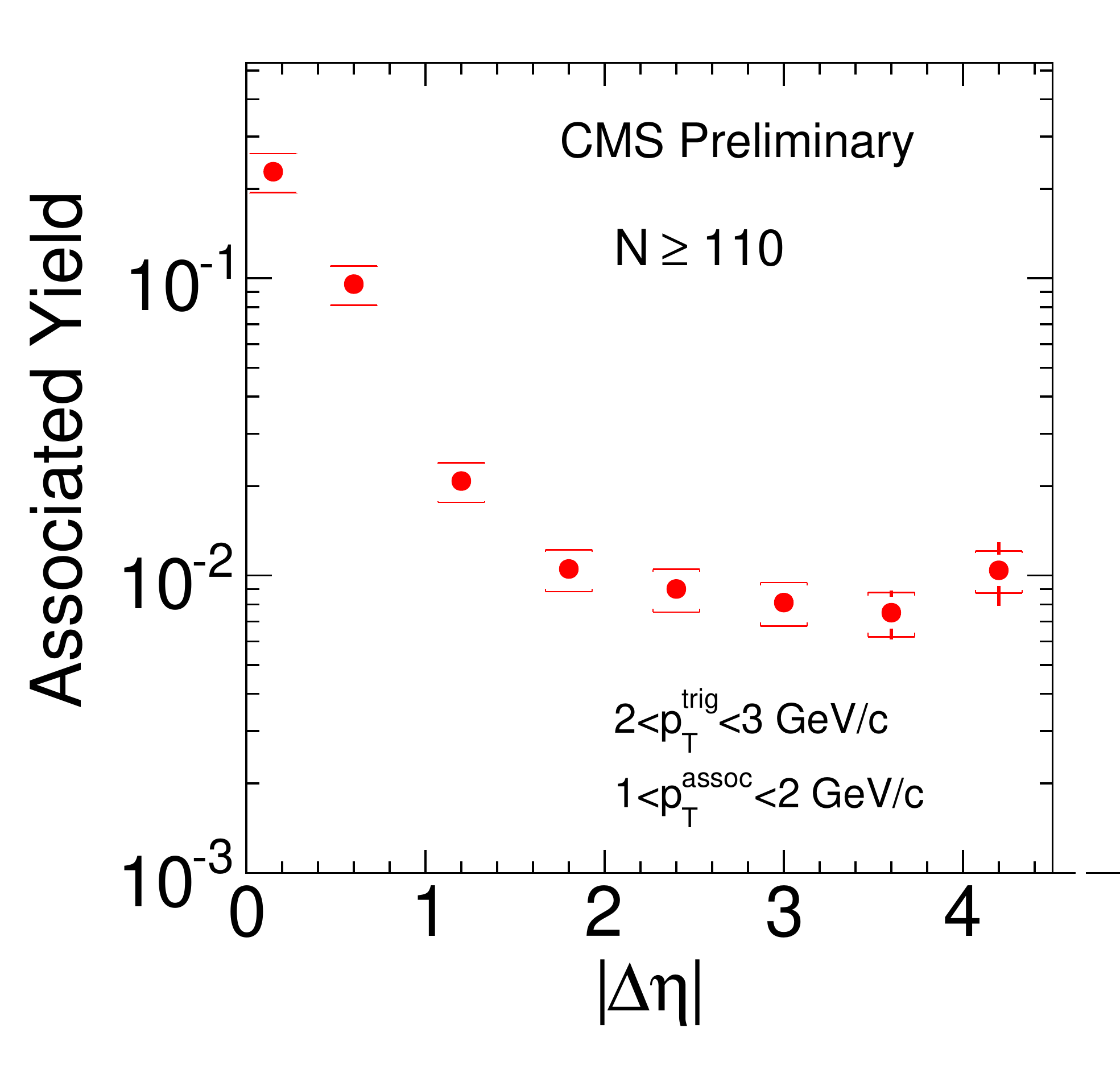}
    \caption{
    Integrated near-side ($|\Delta\phi| $$<$$ \Delta\phi_{\rm ZYAM}$)
    associated yields for $2 $$<$$ \ptttrg $$<$$ 3 \GeVc$ and $1 $$<$$ \pttass $$<$$ 2 \GeVc$,
    above the minimum level found by the ZYAM procedure, as a function 
    of $|\Delta\eta|$ for the high multiplicity ($N \geq 110$) pp 
    collisions at \roots\ = 7~TeV. The statistical uncertainties are 
    shown as bars, while the brackets denote the systematic uncertainties.
    }
    \label{fig:yieldvseta_pp_N110_trg2_ass1}
  \end{center}
\end{figure}

The near-side (small $\Delta\phi$ region) integrated associated 
yield is calculated for both jet and ridge regions relative to 
the constant background, details in Ref.~\cite{ref:HIN-11-001-PAS}. 
Fig.~\ref{fig:yieldvseta_pp_N110_trg2_ass1} 
presents the resulting near-side associated yield as a function of $|\Delta\eta|$ 
(in slices of 0.6 units) in high multiplicity ($N \geq 110$) 
pp collisions at \roots\ = 7~TeV with trigger particles with $2$$<$$\ptttrg$$<$$3 \GeVc$ and 
associated particles with $1$$<$$\pttass$$<$$2 \GeVc$. The high multiplicity data exhibit 
a jet-like correlation peak in the yield for small $|\Delta\eta|$ and show significant 
 and roughly constant yield out to the highest $|\Delta\eta|$ regions.
This is qualitatively similar to what has been observed in central PbPb collisions 
at \rootsNN\ = 2.76~TeV~\cite{ref:HIN-11-001-PAS} but is completely absent in 
minimum bias pp collisions as well as pp MC models.

\begin{figure}[thb]
  \begin{center}
    \hspace{-1.2cm} 
    \includegraphics[width=1.05\textwidth]{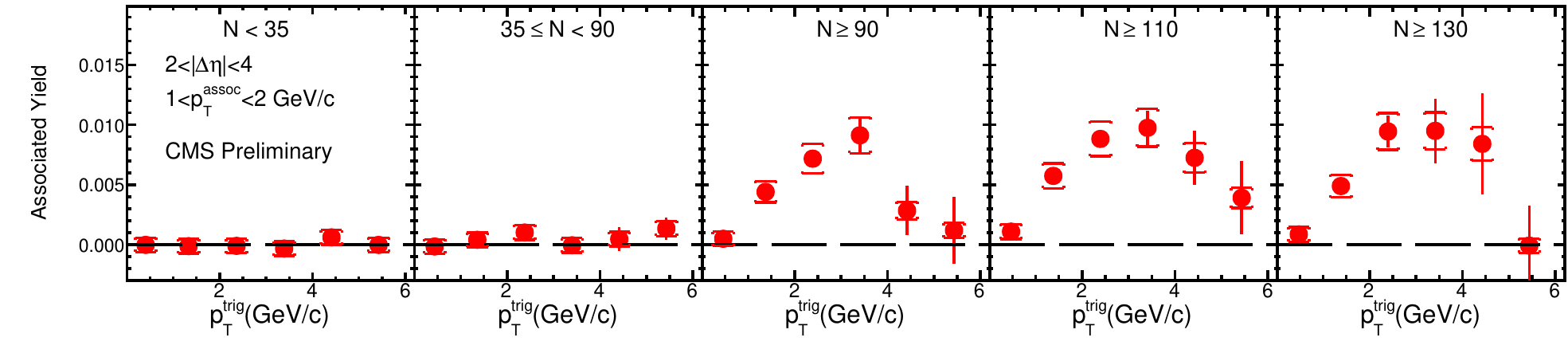}
    \caption{
        Integrated near-side ($|\Delta\phi| $$<$$ \Delta\phi_{\rm ZYAM}$)
        associated yields for the long-range ridge region ($2<|\Delta\eta|<4$) 
        with $1 $$<$$ \pttass $$<$$ 2 \GeVc$,
        above the minimum level found by the ZYAM procedure, as a function 
        of \ptttrg\ for five multiplicity bins ($2 \leq N < 35$, $35 \leq N < 90$,      
        $N \geq 90$, $N \geq 110$, $N \geq 130$) of pp collisions at \roots\ = 7~TeV. 
        The statistical uncertainties are shown as bars, while the brackets 
        denote the systematic uncertainties.
    }
    \label{fig:yieldvspt_pp_ass1_eta24}
  \end{center}
\end{figure}

Figure \ref{fig:yieldvspt_pp_ass1_eta24}
shows the integrated near-side associated yield of the ridge region
correlations with $1$$<$$\pttass$$<$$2 \GeVc$ (the \ptt\ range where the ridge effect 
appears to be strongest) as a function of \ptttrg\ in five bins of event multiplicity.
The ridge yield is almost 
zero for the first two low multiplicity bins in Fig.~\ref{fig:yieldvspt_pp_ass1_eta24}. 
In the high multiplicity region ($N \geq 90$), the ridge yield first increases  
steadily with \ptttrg, reaches a maximum around \ptttrg\ $\sim$ 2--3 GeV/c and drops at higher \ptttrg\ .

\begin{figure}[thb]
  \begin{center}
    \hspace{-0.5cm}
    \includegraphics[width=0.50\linewidth]{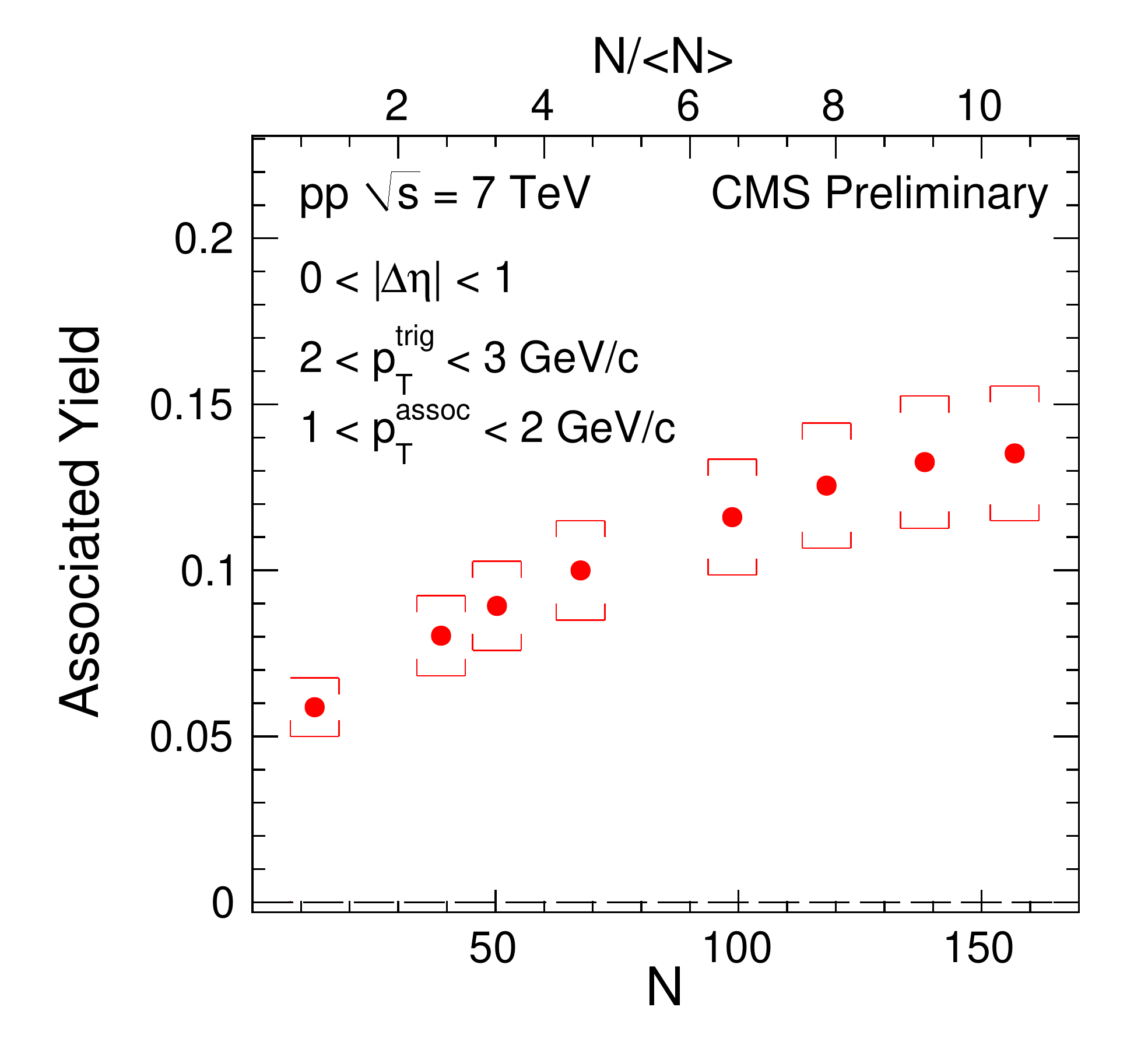} 
    \hspace{-0.2cm}    
    \includegraphics[width=0.50\linewidth]{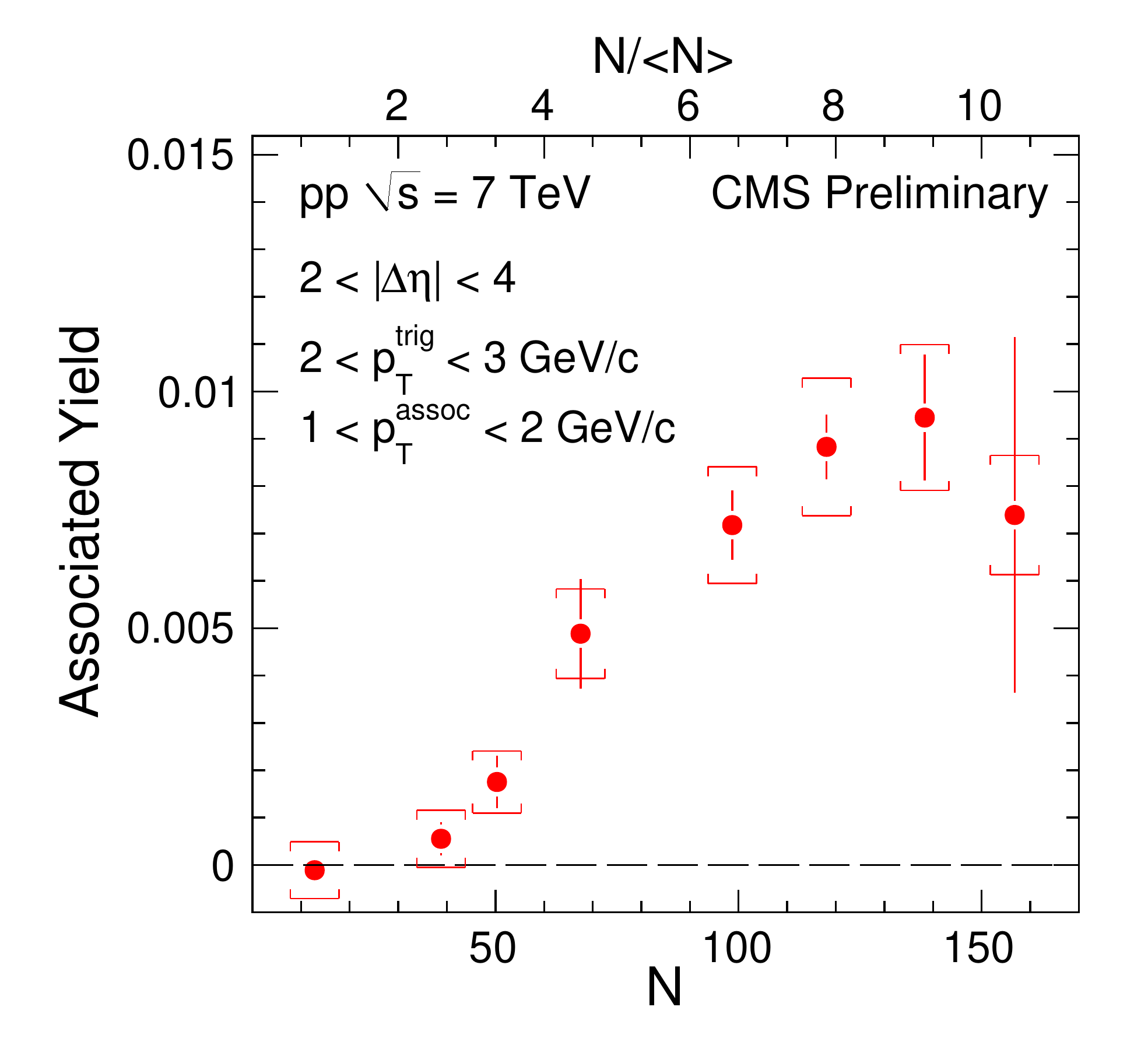} 
    \caption{
        Integrated near-side ($|\Delta\phi| $$<$$ \Delta\phi_{\rm ZYAM}$)
        associated yields for the short-range jet region ($0<|\Delta\eta|<1$) and 
        the long-range ridge region ($2<|\Delta\eta|<4$),
        with $2$$<$$\ptttrg$$<$$3 \GeVc$ and $1 $$<$$\pttass$$<$$ 2 \GeVc$,
        above the minimum level found by the ZYAM procedure, as a function 
        of event multiplicity from pp collisions at \roots\ = 7~TeV. 
        The statistical uncertainties are shown as bars, while the brackets 
        denote the systematic uncertainties.
    }
    \label{fig:yieldvsmult_pp_trg2_ass1}
  \end{center}
\end{figure}

The multiplicity dependence of the near-side associated yield in the jet and ridge region
is illustrated in Fig.~\ref{fig:yieldvsmult_pp_trg2_ass1} for one 
transverse momentum bin of $2$$<$$\ptttrg$$<$$3 \GeVc$ and 
$1$$<$$\pttass$$<$$2 \GeVc$, the \ptt\ bin where the ridge effect appears to be strongest.
The ridge effect gradually turns on with event multiplicity around $N \sim$ 50--60 and 
shows a tendency to saturate when it reaches $N \sim 120$, although this is not yet 
conclusive with current statistical and systematic uncertainties.

In summary, comprehensive studies of the ridge correlation structure in high 
multiplicity pp events, as a function of event multiplicity, particle 
transverse momentum and $\Delta\eta$ using full statistics pp data in 2010, 
are presented in this paper, which provides further 
information on the properties of the 
novel ridge phenomena. 
This is an important step forward toward understanding the physical origin of
the ridge in high multiplicity pp collisions and put additional constraints on various
theoretical models.

\section*{References}

\vspace{-0.5cm}
\providecommand{\href}[2]{#2}\begingroup\raggedright\endgroup

\end{document}